# Multifunctional properties of composition graded Al wires

*Cui Yang[1], Nicolas Masquellier[2], Camille Gandiolle[3], Xavier Sauvage[1*]*

1- Normandie Université, UNIROUEN, INSA Rouen, CNRS, Groupe de Physique des Matériaux, 76000 Rouen, France
2- Nexans, 460 Avenue Durocher Montreal Est, QC H1B 5H6, CANADA
3- Université Paris-Saclay, CentraleSupélec, CNRS, Laboratoire de Mécanique des Sols, Structures et Matériaux, 91190, Gif-sur-Yvette, France

* Corresponding author: xavier.sauvage@univ-rouen.fr

**Abstract**

The potentiality of composition graded AlMgSi wires for optimized combination of electrical conductivity and torsion strength has been investigated. Composition graded wires were obtained by co-drawing commercially pure Al with an AlMgSi alloy followed by diffusion annealing. Diffusion gradients and local hardening response to precipitation treatments were evaluated thanks to nano-indentation measurements. Resulting microstructures with spatial gradients of nanoscaled precipitates were characterized by transmission electron microscopy. Finally, it is shown that such graded structures give rise to an improved combination of electrical conductivity and mechanical strength in torsion as compared to the predictions based on a classical rule of mixture.

Architectured and graded materials have attracted a wide interest since they offer new opportunities for the optimization or combination of properties [1, 2]. Metallic alloys can be architectured or graded by tuning the local composition or microstructural features. It has been shown for example that surface processing by mechanical treatments may successfully produce gradients of structural defects (dislocations, grain boundaries or twins) and internal stresses that positively affect fatigue resistance [3, 4] or evade the strength-ductility trade-off [5-9]. Chemical gradients provide a larger playground to tune locally microstructures and combine properties. They can be created by bottom-up techniques, such as vapor deposition [10], additive manufacturing [11, 12] or electrodeposition [13], or by top-down processes like solid state diffusion [14] or oxidation [2].

Materials for electrical conductors that have to sustain mechanical loads are designed so that the trade-off between strength and electrical conductivity is optimized for the envisioned application. When the weight is a critical factor, aluminum conductors are usually preferred even if they exhibit a maximum conductivity of only 62% of the International Annealed Copper Standard (IACS). For high strength applications, the best combination of properties is usually achieved with alloys from the AlMgSi system, like the 6101 and 6201 industrial alloys [15, 16]. However, for applications where the main solicitation is fatigue in torsion or bending (like medium-voltage loop cables of windmills or wires for industrial robots), graded wires with a high Mg and Si content in the outer part (for maximum local strength) and low Mg and Si content in the core (for a maximum local conductivity) should help to optimize the wire performances. Thus, the aim of this work was to investigate the potentiality of composition graded AlMgSi wires for optimized combination of electrical conductivity and torsion strength.

Composition graded AlMgSi wires were processed in two steps (Fig. 1). At first, a rod (diameter 10mm) of 1370 commercial aluminum alloy was inserted in a tube of 6201 commercial aluminum (outer diameter 16 mm). Before assembly, materials were sand blasted and cleaned with ethanol. The nominal composition of these alloys is given Table 1. Then, this assembly was cold drawn at a drawing speed of 0.2 m/s with diameter reductions ranging from 10 to 20% per pass. Intermediate annealing treatments of 1h at 550°C have been performed at a current diameter of 10 and 4mm to recover ductility and allow further plastic deformation to reach a final diameter of 1mm. In a second step, composition gradients were tuned by thermal diffusion at 550°C (duration up to 50h). This temperature corresponding also to the solution treatment temperature of the 6201 alloy, materials were water quenched at the end of this treatment to keep Mg and Si in solid solution. Then, an aging treatment was performed at 190°C during 4h to achieve precipitation.

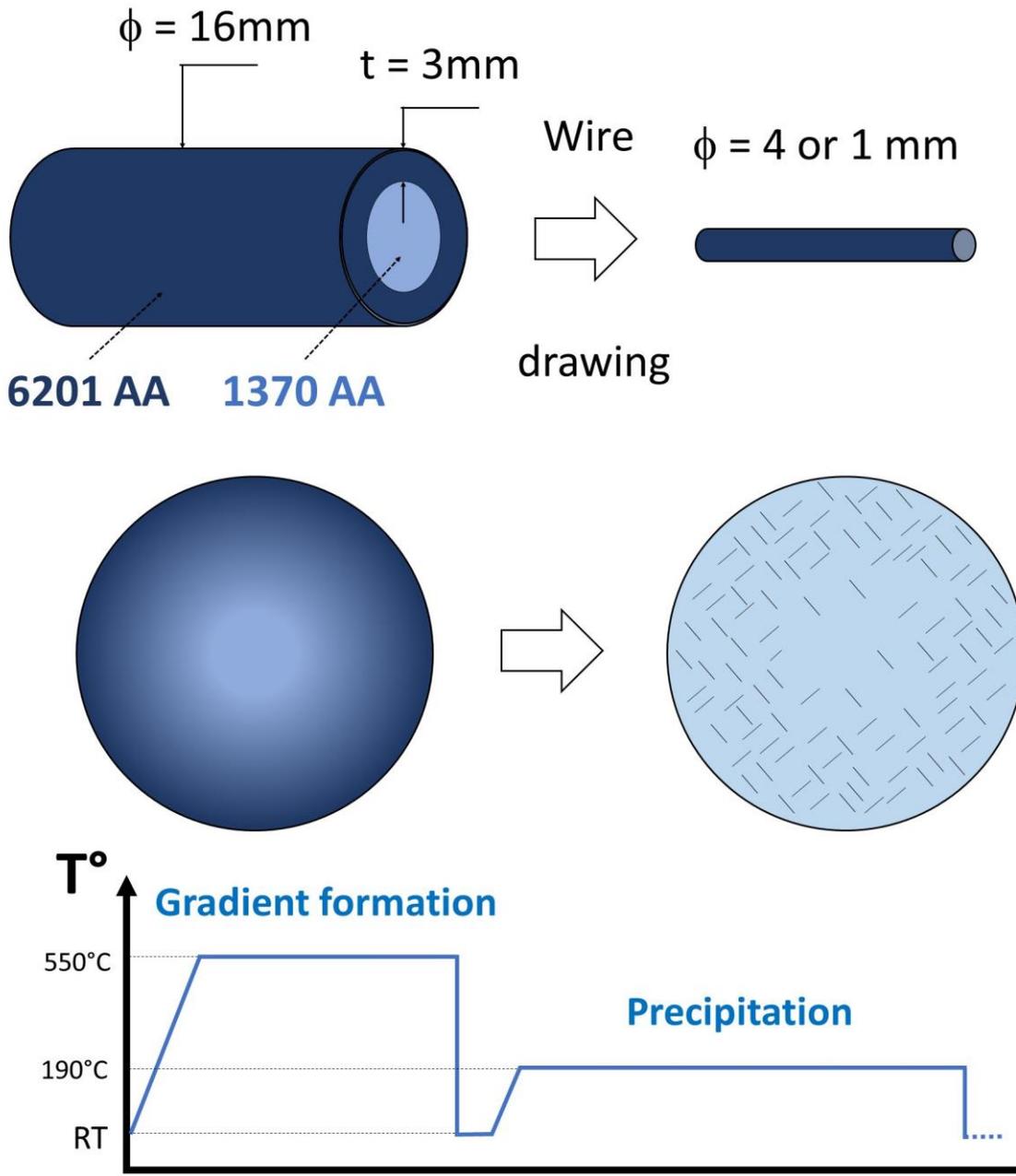

*Figure 1: Schematic representation of the processing route designed to achieve composition graded Al-Mg-Si wires (see text for details).*

The gradient of mechanical properties was evaluated thanks to nano-indentation measurements using a TriboIndenter Hysitron TI 950 with a Berkovitch indenter and a penetration depth of 500 nm (corresponding to an indent size of about 4 µm). The global mechanical strength and conductivity were then assessed. Tensile properties were evaluated with a gauge length of 8mm at a constant cross-head speed of 0.01mm/s. The torsion behavior was evaluated on a torsion bench with a rotation speed of 60rev/min on a 0.265m long sample under a tensile stress corresponding to 25% of the yield stress. The torque and length variations were measured as a function of the number of revolutions. Electrical properties were measured at 20°C on wires with a length of 0.5m with a four-point probe method. Microstructures and especially nanoscaled precipitates were characterized by Transmission Electron Microscopy (TEM). Samples were sliced in the cross-section of wires and electron transparency was achieved thanks to standard electropolishing procedures followed by ion milling using a Precision Ion Polishing System (PIPS, Gatan) at a voltage of 3 kV. Observations were carried out in a JEOL-ARM 200F microscope operated at 200kV. Si and Mg concentration in the wire core were also evaluated thanks to Atom Probe Tomography (APT). Samples were prepared by standard electropolishing procedures and analyzed using a LEAP 4000HR at a temperature of 30K, with electric pulsing (20% pulse fraction, 200kHz).

To validate the experimental approach and to determine optimal diffusion times required to establish composition gradients, solute diffusion and precipitation were investigated in 4mm diameter wires (Fig. 2). Since dislocations introduced during the drawing process also affects the hardness, a short annealing treatment of one hour at 550°C was applied to recover these defects and achieve a reference state. The resulting nano-hardness gradient is relatively sharp (Fig. 2a) and it significantly extends after 50h at 550°C. Solute elements in aluminum enhance the yield stress due to interactions with mobile dislocations. The yield stress increment $\Delta\sigma$ is a function of the solute concentration and can be written as [17, 18]:

$$\Delta\sigma = \sum_i k_i C_i^{2/3} \qquad (1)$$

Where $k_i$ are constants and $C_i$ the concentrations of solutes $i$. Thus, the hardness being proportional to the yield stress (Tabor's law [20]), hardness profiles measured before the precipitation treatment may provide a qualitative estimation of solute concentration profiles in the composition graded wires.

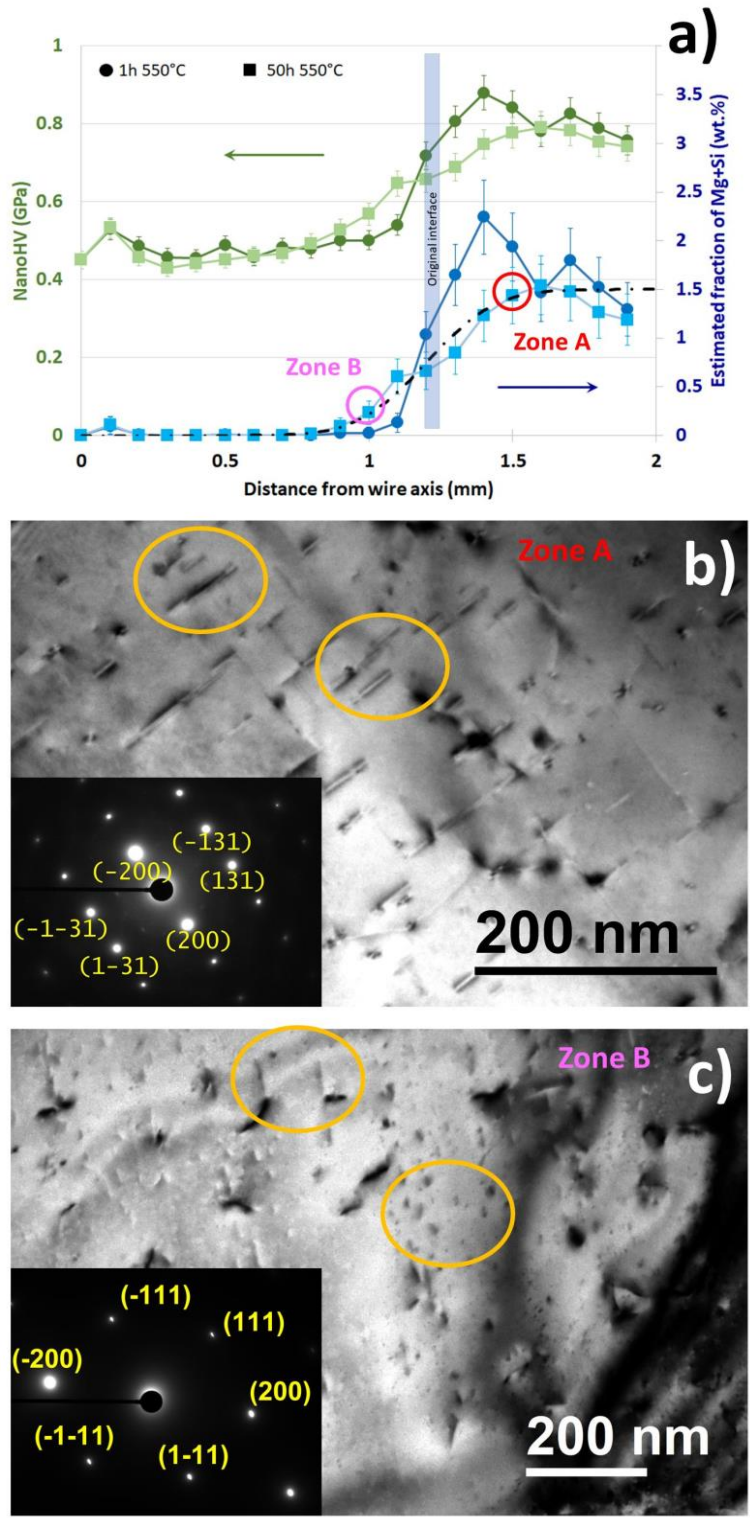

*Figure 2: Figure 2:* (a) Nanohardness profiles (green) in the cross section of the 4mm wire after 1h (circles) and 50h (squares) annealing at 550°C, and corresponding concentration gradients (blue) estimated from Eq. (2) – see text for details. The black dotted line is a fit done with Eq. (3) with $C^0 = 1.5$ at.%, t=50h and D= $10^{-9}$ $cm^2.s^{-1}$ ; TEM bright field images (SAED patterns insets) showing a relatively high density of needle shape nanoscaled precipitates (circled) in Zone A located at a distance of 1.5mm of the wire axis (b) and significantly less in Zone B located at 1mm (c), after 50h annealing at 550°C and 4h aging at 190°C.

Si and Mg in solid solution do not contribute equally to yield stress increment [17]. However, the Mg/Si ratio should not change much along the gradient since these elements exhibit similar diffusion coefficients at 550°C [19]. Then, an average contribution $k^*$ might be considered and assuming proportionality between hardness and yield stress increment (Tabor's law [20]), the hardness increment $\Delta H$ might be written from equation (1) as:

$$\Delta H = k^* C_{Si+Mg}^{2/3} \qquad (2)$$

Where $C_{Si+Mg}$ is the Mg plus Si concentration (wt.%). After one hour at 550°C, the nanohardness difference between the two plateaus corresponding to the 1350 and 6201 alloys is about 300MPa. The solute concentration difference between these two alloys being of about 1.5wt.% (see Table 1), it leads to k*~230 MPa wt%$^{-2/3}$.

Solute concentration profiles displayed on Fig. 2a were estimated thanks to equation (2). After 50h annealing, at 1.5mm from the wire axis, the amount of Mg and Si is still relatively high, between 1 and 1.5 wt.%, while it stands below 0.5 wt.% at a distance of 1mm. Since, the diffusion coefficients of Mg and Si are relatively close at 550°C [19], they can be considered as a unique solute to estimate the composition gradient from the Fick's second law written for a semi-infinite diffusion couple [21]:

$$C_{Si+Mg}(x) = \frac{C_{Si+Mg}^0}{2}\left[1 + \text{erf}(\frac{x}{2\sqrt{Dt}})\right] \qquad (3)$$

Where $C^0_{Si+Mg}$ is the initial Si+Mg concentration of the 6201 alloy, $D$ the mean diffusion coefficient, $t$ the time and $x$ the distance from the initial interface. The composition profile estimated after 50h at 550°C on Fig. 2a was fitted with this equation with a diffusion coefficient D = 10$^{-9}$ cm$^2$.s$^{-1}$ (black dotted line). This is significantly less than diffusion coefficients reported for Mg or Si in Al (at 550°C, they are relatively similar for these two elements, being respectively 8. 10$^{-8}$ and 10$^{-7}$ cm$^2$ s$^{-1}$ [19]). This difference could be due to interactions between Mg and Si that would reduce their mobility. It may also be the result of a not perfectly bonded interface between the 6201 and the 1350 Al alloys during the co-drawing process. Anyway, this mean diffusion coefficient allowed the determination of an optimal annealing time for 1mm diameter wires. To achieve a composition gradient spreading over one third of the radius, two hours at 550°C are necessary.

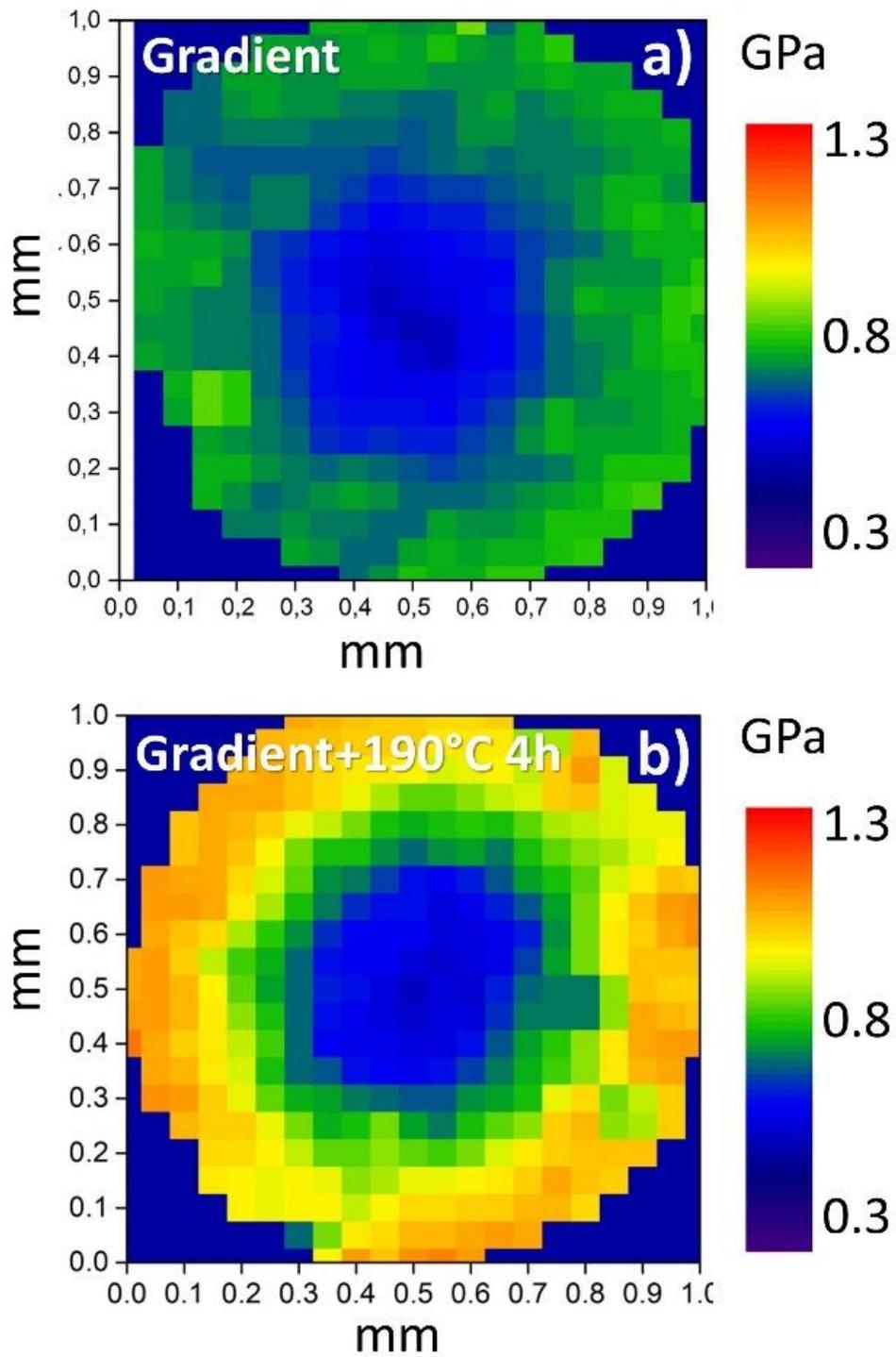

*Figure 3:* *Nanohardness maps of the cross section of the 1mm wire after 2h annealing at 550°C to achieve a composition gradient (a) and after subsequent aging at 190°C during 4h to achieve precipitation hardening (b).*

Once diffusion gradients are established, a classical precipitation treatment of 4h at 190°C can be applied to achieve mechanical strengthening like in standard 6201 alloys. The large difference in Mg and Si concentrations along the gradient naturally leads to very different nanoscaled precipitates [22, 23]. At a distance of about 1.5mm from the axis of a 4mm-wire, a relatively high density of needle shape nanoscaled precipitates was indeed clearly revealed using bright field TEM (Fig. 2b). While shifting toward the wire axis, they progressively disappear: At 1mm (Fig. 2c), a much lower number density of nanoscaled particles could be seen and most of them are not needle shape like the classical β' phase [22]. The hardening resulting from the nucleation of theses precipitates is exhibited on the nano-hardness maps of the 1mm wires (Fig. 3). The original axisymmetric structure of the composite created with the 1350 alloy rod inserted in the 6201 alloy tube was relatively well preserved during the drawing process. In the precipitated state (Fig. 3b), the gradient of properties is nicely exhibited on the map, with a nano-hardness ranging from a mean value of 1.2 GPa at the wire edge, down to 0.4 GPa at the center. Thus, as expected, a strong hardening occurred due to precipitation [17, 24] in the solute rich outer part of the wire. The core (about 0.5mm in diameter) did not significant harden because only a very small fraction of solutes has migrated into this region during the diffusion annealing treatment. APT analyzes carried out in the wire core region revealed indeed that as compared to the original 1350 alloy, the amounts of Si and Mg have increased of only about 0.1 and 0.01wt.% respectively (see Table 1).

The tensile behavior of this composition graded wire was compared to the conventional 6201 and 1350 aluminum alloys (Fig. 4a and Table 1). Before the precipitation treatment, when only a solute concentration gradient was created by annealing, the yield stress is slightly higher than the 1350 alloy (30 MPa against 15MPa), but the strain hardening rate is much larger, giving rise to a significantly larger Ultimate Tensile Stress (UTS) (90 MPa against 40 MPa). This is the result of the interaction between moving dislocations and solute elements that typically gives rise to highest dislocation densities. After the precipitation treatment during 4h at 190°C, the yield stress dramatically increases (up to 115MPa) but does not reach the yield stress of the 6201 alloy treated in similar conditions (250MPa). The uniform elongation of the graded material is however twice larger than that of the 6201 alloy. This might be attributed to the soft and ductile wire core whose properties were relatively unchanged (Fig. 3) and which might help to delay necking and thus damage propagation.

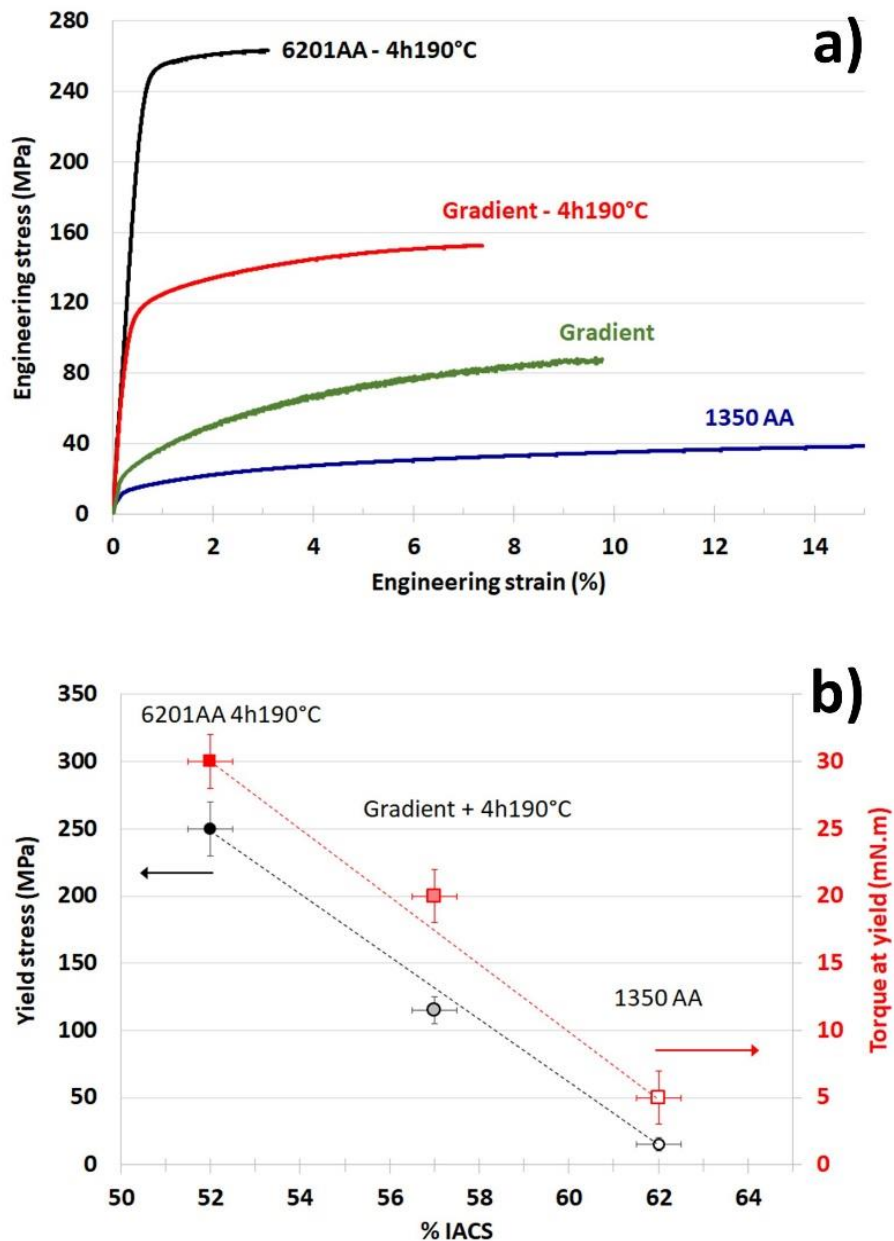

*Figure 4:* (a) Strain stress plots of 1mm wires of 6201AA (homogenized and aged 4h at 190°C), 1350AA (annealed at 550°C), composition graded material (annealing 2h at 550°C followed by quenching), and composition graded material after precipitation treatment at 190°C during 4h. (b) Yield stress and torque at yield versus the electrical conductivity of the wires. Doted lines correspond to extrapolations based on a linear rule of mixture between the two original alloys.

As shown on Fig. 4b, the combination of yield stress and electrical conductivity of the composition graded wire stands close to the prediction based on a linear rule of mixture and on properties of the original 1350 and 6201 alloys. Thus, the gain in electrical conductivity obtained thanks to the low Mg and Si concentration in the wire core is fully balanced by the drop of yield stress. The combination of properties is even slightly below the rule of mixture prediction, probably because of the small amount of Mg and Si that have migrated in the core region of the wire. Indeed, the contribution of Si and Mg in solid solution to the electrical resistivity of aluminum is relatively similar (about 5 $10^{-7}\Omega$cm at.%$^{-1}$ [18, 25]), and an increase of 0.1at.% of solute as measured by APT, would shift the conductivity of the wire core by about -1%IACS. However, the torsion behavior evaluated thanks to the torque at yield in torsion shows that the graded materials stands above the prediction based on a linear rule of mixture with a positive shift of about 1%IACS (Fig. 4b). This is obviously the result of the hardness gradient displayed on Fig. 3b, since a wire under torsion stress has to sustain the maximum load in the skin while the shear stress linearly decreases down to zero on the axis. Besides, like under tensile stress, a larger uniform plastic deformation (20% more) could be achieved for the graded wire under torsion as compared to the 6201 alloy (Table 1).

In conclusion, this work demonstrates the potentiality of composition graded aluminum wires for applications where an optimal combination of electrical conductivity and mechanical strength in torsion is required. It is envisioned that tuning the Mg and Si contents of the starting outer AlMgSi alloy (especially using higher concentrations than in the 6201 alloy of the present study) could open a broad panel of Al wires with optimized multifunctional properties.

**Acknowledgements**

Prof. Jean-Hubert Schmitt is gratefully acknowledged for fruitful discussions.

**Tables**

|  | 6201AA (4h190°C) | 1350AA | Composition graded (4h 190°C) |
|---|---|---|---|
| wt.% Mg | 0.6 – 0.9 | --- | *at wire center* 0.01±0.008 |
| wt. % Si | 0.5 – 0.9 | < 0.1 | *at wire center* 0.2±0.001 |
| Yield stress (MPa) | 250 ± 20 | 15 ± 5 | 115 ± 10 |
| UTS (MPa) | 260 ± 20 | 40 ± 5 | 150 ± 10 |
| Uniform elongation (%) | 3.5 ± 0.5 | 20 ± 5 | 7 ± 1 |
| Torque at yield (mN.m) | 25 ± 5 | 7 ± 1 | 15 ± 5 |
| Maximum torque (mN.m) | 40 ± 10 | 17 ± 2 | 30 ± 5 |
| Torsion at torque max (rev/m) | 100 ± 10 | 800 ± 80 | 120 ± 10 |
| Electrical conductivity %IACS | 52 ± 0.5 | 62 ± 0.5 | 56.5 ± 0.5 |

*Table 1: Tensile, torsion and electrical experimental data of the two reference materials (6201 AA homogenized and aged 4h at 190°C, 1350AA annealed at 550°C) and of the 1mm composition graded wire (annealed 2h at 550°C, quenched and aged at 190°C during 4h). The yield stress is the conventional engineering stress measured at 0.2% strain and UTS stands for Ultimate Tensile Stress. The conductivity is given in % of international annealed copper standard (IACS). Si and Mg concentrations are nominal compositions for 6201 and 1350 aluminum alloys and composition measured by APT at the center (±0.1mm) of the 1mm composition graded wire.*